\documentclass[onecolumn,showpacs,showkeys,preprintnumbers,amsmath,amssymb]{revtex4}

\usepackage{graphicx}
\usepackage{dcolumn}
\usepackage{bm}
\usepackage{epsfig}
\begin{document}

\preprint{IST/IPNF 2008-Martins Pinheiro}

\title[]{Fluidic electrodynamics: Approach to electromagnetic propulsion}

\author{Alexandre A. Martins}\email{aam@ist.utl.pt}
\address{Institute for Plasmas and Nuclear Fusion \& Instituto Superior T\'{e}cnico, Av. Rovisco Pais, 1049-001 Lisboa,
Portugal}

\author{Mario J. Pinheiro} \email{mpinheiro@ist.utl.pt}
\affiliation{Department of Physics and Institute for Plasmas and
Nuclear Fusion, Instituto Superior T\'{e}cnico, Av. Rovisco Pais,
1049-001 Lisboa, Portugal}

\thanks{The authors gratefully acknowledge partial financial support by the
Funda\c{c}\~{a}o para a Ci\^{e}ncia e a Tecnologia (FCT). We would
also like to thank important financial support to one of us (A.A.M.)
in the form of a PhD Scholarship from FCT.}




\date{\today}

\begin{abstract}
We report on a new methodological approach to electrodynamics based
on a fluidic viewpoint. We develop a systematic approach
establishing analogies between physical magnitudes and isomorphism (structure-preserving mappings)
between systems of equations. This methodological approach allows us to
give a general expression for the hydromotive force, thus
re-obtaining the Navier-Stokes equation departing from the
appropriate electromotive force. From this ground we offer a fluidic
approach to different kinds of issues with interest in propulsion,
e.g., the force exerted by a charged particle on a body carrying
current; the magnetic force between two parallel currents; the
Magnus's force. It is shown how the intermingle between the fluid
vector fields and electromagnetic fields leads to new insights on
their dynamics. The new concepts introduced in this work suggest
possible applications to electromagnetic (EM) propulsion devices and
the mastery of the principles of producing electric fields of
required configuration in plasma medium.
\end{abstract}
\maketitle

\section{Introduction}

The analogy between the vector fields of electromagnetism and the
hydrodynamic fields provides far-reaching insights into
electromagnetic (EM) fields~\cite{Rousseaux_02,Liu_93,Sridhar_98,MartPin_08}
and it has been used to solve complex problems, like the turbulent
fluid flow by Marmanis~\cite{Marmanis_98}.

Several steps have been made in this direction. P. Holland~\cite{Holland_05} deduced the set of
Maxwell's equations from continuum mechanics by generalizing the
spin-0 theory to a general Riemannian manifold. On a different
ground, the conceptual similarities between condensed matter and the
quantum vacuum allows one to simulate phenomena in high-energy
physics and cosmology using quantum liquids, Bose-Einstein
condensates and all scenarios appearing in condensed matter
systems~\cite{Volovik}. The failure to develop a quantum field
theory of gravitation leads to a relativistic description of
rotating space-time that reveals striking similarities between the
vacuum state with quantum
condensate~\cite{Mazur_05,Vinem,Volovik_98}. Meholic and Fronning~\cite{Meholic} proposed the concept of a superluminal space inside which a vessel would be propelled through the ``superlight" continuum by a field that interacts with static electromagnetic forces.

A fluid-dynamic approach to the relativistic electromagnetic field has been done by Kaufman~\cite{Kaufman1,Kaufman2}, treating the electromagnetic fluid as a ``fluid" with a ``definite and calculable velocity relative to an observer"~\cite{Kaufman2} and taking into account an additional flow-work term, he obtained a Lorentz-invariant electron mass. We have obtained similar conclusions through a different procedure, using the convective derivative of the EM ``fluid"~\cite{Pinheiro1,AlexPin1,AlexPin2}.

These similarities between vast areas of physics undoubtedly constitute a successful
example of the unity of physics and illuminates the physical reality
hidden inside the physical vacuum.

In this paper we introduce a ``fluidic electrodynamic" approach to
the electromagnetic fields in terms of the potential functions
($\mathbf{A}$, $\phi$) and their material derivative, as they
emerge in quantum mechanics as more fundamental quantities than the
($\mathbf{E}$, $\mathbf{B}$) fields, predicting certain quantum
interference effects, like the Aharonov-Bohm (AB) effect and the
single-leg electron interferometer effect known as the Josephson
effect.

The ``fluidic electrodynamics" viewpoint brings far-reaching results
as analogies are concerned. In addition, the simplified methodology
offered in this article helps to solve certain problems by anyone
with reasonable knowledge of electrodynamics and to apply to
situations of experimental interest with sufficient accuracy, before
undergo more complicated procedures.

\section{Fluidic approach to electromagnetism}

The founders of electromagnetism envisaged the
``aether" as an ``elastic solid" (e.g., Faraday's ``electro-tonic
state"-mechanical medium subject to certain states of tension and
motion), and Maxwell later on used the representation of Faraday's
line of magnetic force as the velocity of an incompressible
fluid~\cite{Yang_1}, introducing in a quantitative form the
Faraday's law of induction under the form $\mathbf{E}=-\partial
\mathbf{A}/ \partial t$.

Bernhard Riemann attributed the cause of gravitation and light ``in
the form of motion of a substance that is continuously spread
through all infinite space"~\cite{Riemann}. Riemann conceived this
substance as a {\it physical space} whose points move in {\it
geometrical space}. As far as the term ``aether" is introduced hereby, we could argue that it might be better to use it, instead of {\it physical vacuum}, or just {\it vacuum}, since vacuum usually means ``nothing whatsoever", while {\it physical space} confuses the word either with {\it vacuum} or mix up physics with geometry~\cite{Armstrong}.

The special theory of relativity attributed to the ``aether" the
quality of a ``superfluous" artifact to think about natural
phenomena, but currently there is a strong need to rethink this
medium as a true physical vacuum. In fact, Dirac postulated the
existence of an ``aether" needing an ``elaborate mathematics for its
description" ~\cite{Dirac_01} and later he proposed a new classical
theory of electrons formulated in a hamiltonian form and not based
on gauge transformations~\cite{Dirac_51}. According to him, the 4-vector potential
should verify the condition $A_{\mu}A^{\mu}=m^2/e^2$ and the
4-velocity field should be given by $V_{\mu}=eA_{\mu}/m$. The velocity $V$
appears in the vector potential with the physical significance of
{\it the velocity with which an electron charge must flow in the ether}.
Dirac interpreted this velocity field as something real, even in the
absence of electric charges. By the contrary, Poincar\'{e} advocated
that the electromagnetic energy was a fictitious fluid transported
in space according to Poynting's laws - in fact, according to his
own words, a useful ``mathematical fiction"~\cite{Poincare_01}.

Remarkably, the experimental findings by Graham and
Lahoz\cite{Graham_80} implies that the vacuum is the seat of
``something in motion", in the way how Maxwell envisaged the
``aether". More recently, it was shown that a medium in uniform
motion with velocity $v$ plays the role of the vector potential, while
the charge is proportional to Fresnel's dragging coefficient for
light in moving media~\cite{Leonhardt}. In the framework of general
relativity, Keech and Corum~\cite{Corum} have shown that an electric
null current is accompanied by a neutral fluid current, and it
is this null fluid that transports the energy instead of the
electromagnetic field.

Reasoning along this line of thought we thus attribute to the vector
potential the property of the velocity of a ``fluid" embedded in the
physical vacuum. We recall that the physicality of the vector
potential is now well proven experimentally~\cite{Tonomura}. We do
not intend here, however, to describe the inner nature of a so
pervasive and evasive medium, be it a mechanical medium whose
deformations correspond to the electromagnetic fields, or a locally
preferred state of rest. However, it is particularly relevant that
currently there have been advanced some explanations about the origin of
``dark matter," inventing new particles, as the WIMP (weekly
interactive massive particle) or the neutralino (e.g., Ref.~\cite{Markevitch}). Observations of our
Galaxy and other major galaxies help to discover vast coronas
extending far beyond the visible stellar systems. Although they do
not emit visible light, their mass may exceed the total mass of the
stars they surround. This question might be related to the Graham
and Lahoz experimental findings and quoted above, that ``something
in motion" is not took into account. As we know that rotating
gravitational fields can generate electromagnetic
fields~\cite{Teller}, it was recently shown that within the
approximation of the linearized Einstein-Maxwell theory on flat
spacetime, an oscillating electric dipole is the source of a spin-2
field, that is, electromagnetic waves harbour gravitational
waves~\cite{Bramson}, possibly interbreeding each other.

A vector potential vector field is created whenever an electric
charge moves or an electric current is produced by an
electromagnetic system. We do not address here the specific problem
of the motion of a given particle in this ``fluid", problem that is
addressed, e.g., in Ref.~\cite{Dimtryev_01}. We are here interested in
describing the effects produced by inducing a flow of the vector
potential around material bodies.

To describe the electromagnetic field it is necessary to define the
electric field $\mathbf{E} (\mathbf{r},t)$, the magnetic field
$\mathbf{B} (\mathbf{r},t)$; charge density $\rho (\mathbf{r},t)$;
and the charge velocity $\mathbf{v} (\mathbf{r},t)$. If we
have a charged particle with position vector $\mathbf{r_i}
(\mathbf{r},t)$, the charge density is $\rho (\mathbf{r},t)=e
\delta(\mathbf{r}-\mathbf{r_i}(t))$. However, it is more advantageous to associate the set of Maxwell's equations with the electromagnetic potentials
$\mathbf{A}(\mathbf{r},t)$ and $\phi (\mathbf{r},t)$ through the
relationships~\cite{Okun,Pinheiro1,AlexPin1}:
\begin{equation}\label{eq1}
\mathbf{E} (\mathbf{r},t)=-\frac{1}{c}\frac{d \mathbf{A}}{d t} -
\nabla \phi,
\end{equation}
and
\begin{equation}\label{eq2}
\mathbf{B}=[\nabla \times \mathbf{A}].
\end{equation}
Note that we introduced into Eq.~\ref{eq1} the convective derivative $d/dt=\partial / \partial t + \mathbf{v} \cdot \nabla$, instead of the Maxwell-Einstein operator $\partial /\partial t$ (see, e.g., Ref.~\cite{Pinheiro1,AlexPin1,AlexPin2}).

We define in Table~\ref{table1} the correspondence of field
variables in electromagnetism and hydrodynamics. In accordance with ~\cite{Semon,Sivardiere,Guyon,Rousseaux_02}, we make the
analogy of the magnetic induction field with vorticity $\mathbf{\omega}$. We also may notice that the
electromagnetic analogue of one of the three inertial forces that is
observed in a frame rotating about a point with angular velocity
$\Omega$, $\mathbf{F}_3=-m [\dot{\Omega} \times \mathbf{r}]$ is the
Faraday's magnetic induction law~\cite{Semon_81,Dempsey}.

\begin{table}
\caption{\label{table1} Correspondence of field variables in
electromagnetism and hydrodynamics.}
\begin{tabular}{|c|c|}
\hline \hline
Electromagnetism & Hydrodynamics\\
\hline

$\frac{q}{\varepsilon_0}$ & Kraftquelle - $Q$ \\
Permeability of the vacuum - $\mu_0$ & Mass density - $\rho$\\
Electric potential - $\phi(\mathbf{r},t)$ & Massic enthalpy - $\pi/\rho (\mathbf{r},t)$ \\
Scalar potential - $\chi$ & Potential velocity - $\Phi$ \\
Vector potential - $\mathbf{A} (\mathbf{r},t)$ & Velocity (or hydrodynamic momentum) - $\mathbf{u} (\mathbf{r},t)$ \\
Electric field - $\mathbf{E} (\mathbf{r},t)$ & Lamb vector - $\mathbf{l} (\mathbf{r},t)$ \\
Magnetic field - $\mathbf{B} (\mathbf{r},t)$ & Vorticity - $\mathbf{\omega} (\mathbf{r},t)$ \\
Voltage (or electric tension) $U (\mathbf{r},t)$ & $p (\mathbf{r},t)=\rho (\mathbf{r},t) \Phi (\mathbf{r},t)-\frac{\rho (\mathbf{r},t)}{2}u(\mathbf{r},t)^2$ \\
Electric current - $I$ & Circulation - $\Gamma$ \\
Electromotive force - $\mathbf{E}=-\frac{\partial
\mathbf{A}}{\partial t} -\nabla \phi -\nabla(\mathbf{v \cdot \mathbf{A}}) + [\mathbf{v} \times
\mathbf{B}]$
& Hydromotive force - $\mathbf{E_H}=-\frac{\partial \mathbf{u}}{\partial t}-\nabla(\frac{p}{\rho}+\frac{u^2}{2})-[\mathbf{\omega} \times \mathbf{u}]$\\
\hline \hline
\end{tabular}
\end{table}

Table~\ref{table2} shows the equivalent equations as they are known in
electrodynamics and hydrodynamics. $n(\mathbf{r},t)$ denotes the
``hydrodynamic" charge density~\cite{Marmanis_98,Rousseaux_07}.

We recall here that the vorticity field is defined by:
\begin{equation}\label{}
\mathbf{\omega} = \nabla \times  \mathbf{u},
\end{equation}
and it obeys to the causal relationships:
\begin{equation}\label{}
\nabla \cdot \mathbf{\omega}=0,
\end{equation}
and,
\begin{equation}\label{}
\frac{\partial \mathbf{\omega}}{\partial t} = - \nabla \times
\mathbf{l}.
\end{equation}
Here, $\mathbf{l}$ defines the Lamb vector,
$\mathbf{l}(\mathbf{r},t)=[\mathbf{\omega} \times \mathbf{u}]$. This
mathematical relationship suggests that the physical entity
we call magnetic field is created by ``something" in rotational
motion. This idea is embedded in the Maxwell's vortex mechanical
model of the electromagnetic ether that depicts spinning vortices
representing a magnetic field, while the lateral motion of the
smaller idle wheel particles would represent the electric
current~\cite{Hunt}. The concept of Lamb vector and hydrodynamic charge have been shown to be useful in locating and characterizing vortex structures and turbulence~\cite{Rousseaux_07}. The term $\mathbf{I}$ inside the Lamb vector governing equation is the turbulent current vector~\cite{Marmanis_phd}.

We may notice here that the vector potential $\mathbf{A}$ in magnetostatics is
quite analogous to the potential $\phi$, (and it has the same mathematical
solutions) in electrostatics (e.g., Ref.~\cite{Schwartz}).

\begin{table*}
\caption{\label{table2} Equivalent equations in electrodynamics and
hydrodynamics.}
\begin{ruledtabular}
\begin{tabular}{ccc}
  $\nabla \cdot \mathbf{B}=0$ & Thomson's equation & $\nabla \cdot \mathbf{\omega}=0$ \\
  $\frac{\partial \mathbf{B}}{\partial t}=-[\nabla \times \mathbf{E}]$ & Faraday's equation & $\frac{\partial \mathbf{\omega}}{\partial t}=-[\nabla \times \mathbf{l}]$ \\
  $\nabla \cdot \mathbf{E}=\rho/\varepsilon_0$ & Gauss's equation & $\nabla \cdot \mathbf{l}=-\nabla^2 \Phi=n(\mathbf{r},t)$ \\
  $\frac{\partial \mathbf{E}}{\partial t}=c^2 [\nabla \times \mathbf{B}]- \mathbf{J}$ & Amp\`{e}re's equation & $\frac{\partial \mathbf{l}}{\partial t}=c^2 [\nabla \times \mathbf{\omega}]- \mathbf{I}$ \\
\end{tabular}
\end{ruledtabular}
\end{table*}

Recall that the hydrodynamic circulation is given by:
\begin{equation}\label{eq3}
\Gamma_H =\int_{\gamma} (\mathbf{u} \cdot d \mathbf{s})=\int\int_S
([\nabla \times \mathbf{u}] \cdot \mathbf{n})dS = \int \int_S (\omega
\cdot \mathbf{n}) dS,
\end{equation}
while in electromagnetism circulation is given by:
\begin{equation}\label{eq4}
\Gamma_{EM} =\int_{\gamma} (\mathbf{A} \cdot d
\mathbf{s})=\int\int_S ([\nabla \times \mathbf{A}] \cdot
\mathbf{n})dS=\int\int_S (\mathbf{B} \cdot \mathbf{n})dS=\Phi_B.
\end{equation}
Here, $\Gamma$ is the line integral (circulation) around a closed
curve $\gamma$ of the fluid velocity, and $d \mathbf{s}$ is a unit
vector along $\gamma$. Hence, $\Gamma$ is the analogue of the
electric current, but it does not presuppose a ``closed circuit". It
is just the circulation around the curve $\gamma$ that delimits the
surface $S$ crossed by the flux of a given ``kind" of substance.
This is an outcome of Stoke's theorem.
This quantity $\Gamma_{EM}$ is in fact the magnetic flux. The
$\mathbf{A}$ field is the vector potential and $\mathbf{B}=[\nabla
\times \mathbf{A}]$. Both Eq.~\ref{eq3}-~\ref{eq4} are applications
of the Kelvin-Stokes theorem relating the surface integral of the
curl of a vector field over a surface $S$ in the Euclidean 3-dimensional space
to the line integral of the vector field along $\gamma$ with a
positive orientation, such that $\mathbf{ds}$ points counterclockwise
when the surface normal $\mathbf{n}$ points toward the viewer.

We have shown in previous publications~\cite{Pinheiro1,AlexPin1,AlexPin2} that the electric field is
given with generality in the form:
\begin{equation}\label{eq5}
\mathbf{E}=-\nabla\phi - \frac{\partial \mathbf{A}}{\partial t} +
[\mathbf{v} \times \mathbf{B}] - \nabla(\mathbf{v} \cdot \mathbf{A}).
\end{equation}
Eq.~\ref{eq5} contains the standard Lorentz force, but also a new extra term, which come out naturally from the analysis of the set of Maxwell's equation in moving electromagnetic systems~\cite{Pinheiro1,AlexPin2,Pinheiro3}, which is $\ \nabla (\mathbf{v} \cdot \mathbf{A})$. We recall that Eq.~\ref{eq5} is an outcome of the use of the convective derivative $d/dt$ inside the Maxwell set of equations. It can be readily shown that $d/dt$ is Galilean invariant whereas the Maxwell-Einstein operator $\partial /\partial t$ is not~\cite{PhippsJr}. Multiplying Eq.~\ref{eq5} by the elementary charge $dq$ it gives the electrodynamic force law. However, it appears now a longitudinal force acting on the elementary charge, $-dq \nabla (\mathbf{v} \cdot \mathbf{A})$. According to Boyer~\cite{TBoyer_06} this force could possibly introduce an EM-lag effect on a beam of electrons accounting for the AB effect. In this case, the AB effect might be a local effect, making the quantum topological interpretation untenable. Recent experiments~\cite{Caprez_07}, however, observe no lag effect for electrons passing a macroscopic solenoid. According to Boyer~\cite{TBoyer071} this experiment does not yet rule out the controversy, since the classical lag effect depends strongly upon the details of the interaction between electrons and the current of the solenoid~\cite{Boyer_73}.

From the above considerations we can set up an isomorphism (structure-preserving mapping) between EM-Field and fluid dynamics - what we call ``fluidic electrodynamics" - obtaining the following expression for an element of hydrodynamic force acting on a given physical system:
\begin{equation}\label{eq5a}
\mathbf{E}_H= -\nabla \Phi - \frac{\partial \mathbf{u}}{\partial t}
+ [\mathbf{u} \times \mathbf{\Omega}] - \nabla u^2.
\end{equation}
From the purely theoretical point of view, Eqs.~\ref{eq5} and ~\ref{eq5a} are completely analogous mathematically, while describing different kinds of ``fluids", all symmetry being recovered.

Using the mathematical identity:
\begin{equation}\label{eq5b}
\frac{1}{2} grad (u^2) = [\mathbf{u} \times rot \mathbf{u}] +
(\mathbf{u} \cdot \nabla) \mathbf{u},
\end{equation}
into the above Eq.~\ref{eq5a}, we obtain after a short calculation
the {\it hydrodynamic motive} force in a suitable form:
\begin{equation}\label{eq5c}
\mathbf{E}_H=-\nabla \left(\frac{p}{\rho} + \frac{1}{2}u^2 \right) -
\frac{d\mathbf{u}}{dt} - [\mathbf{\omega} \times \mathbf{u}].
\end{equation}
Multiplying Eq.~\ref{eq5c} by the elementary mass $dm$ we obtain a
dynamic equation of motion for the elementary particle of mass $dm$:
\begin{equation}\label{eq6}
d \mathbf{F}^{ext} = - dm \mathbf{E}_H = - dm \nabla
\left(\frac{p}{\rho} + \frac{1}{2} u^2 \right) - dm \frac{d
\mathbf{u}}{d t} - dm [\mathbf{\omega} \times \mathbf{u}].
\end{equation}

Hence, the force acting on a moving charge with velocity
$\mathbf{v}$ relatively to a frame at rest, must be given by:
\begin{equation}\label{eq7}
\mathbf{F}=q\mathbf{E}=-q\nabla\phi-q\frac{\partial
\mathbf{A}}{\partial t}+q[\mathbf{v} \times \mathbf{B}] -q
\nabla(\mathbf{v} \cdot \mathbf{A}),
\end{equation}
while the corresponding dynamical equation of hydrodynamics for an
ideal fluid, the Euler-Gromeko equation, is given by:
\begin{equation}\label{eq8}
\rho \frac{\partial  \mathbf{u}}{\partial t} + \mathbf{f}^{ext}=-\nabla
\left(p+\frac{\rho}{2}u^2 \right) - \rho [\mathbf{\omega} \times
\mathbf{u}],
\end{equation}
with $\mathbf{f}^{ext}$ representing an external force per unit of
mass (in SI units N$/$m$^3$) - a source term. This equation has the
advantage to appear with the angular velocity, and as it happens
with Euler equation, Eq.~\ref{eq8} obeys to the same kind of initial
conditions and boundary conditions. It is worth to recall the
contributions of two different kind of forces to the fluid dynamics,
the gradient of enthalpy $h = p + \rho u^2/2$ and the last term on the
right-hand-side, which describes strain condition of the medium. Eq.~\ref{eq8} reproduces the dynamics of an incompressible fluid, but it has been shown by Murad~\cite{Murad_06} that classical incompressible steady-state subsonic flow solutions can be transformed into viscous solutions by changing the definition of the potential.

The last term of Eq.~\ref{eq7} represents a new term and it might be important where there is chaotic motion, like in arcs and plasmas~\cite{Pinheiro3}, or in special geometries. Others authors argued the need of another term correcting Lorentz force, among them, Thomas Phipps~\cite{ESJ_97}, P. Graneau~\cite{Graneau_96}, Cavalleri {\it et al.}~\cite{Hadronic_98}, Monstein~\cite{Monstein_98}, but they have not given a consistent formulation as we do here.

\subsection{Euler and Bernoulli's integrals of the fluid dynamic equations}

Both the Euler equation and the Euler-Gromeko equation can be
integrated in special cases provided certain conditions are given.
Here, we follow the presentation given in standard
textbooks (e.g., Ref.~\cite{Faddeev}). Let us consider again
Eq.~\ref{eq7}. Assuming that our physical system at study is
submitted to an external force of the type $\mathbf{f}^{ext}=-\rho_c
\mathbf{E}$, under the assumption of constant charge density, with
$\rho_c=e(n_e-Zn_i)$ denoting the charge density for a plasma fluid,
we can rearrange Eq.~\ref{eq7} in the following form :
\begin{equation}\label{eq12a}
\rho_c \frac{\partial \mathbf{A}}{\partial t} = -\nabla \left[\rho_c
(\mathbf{v} \cdot \mathbf{A}) + \rho_c \phi \right] + \rho_c
[\mathbf{v} \times \mathbf{B}] - \mathbf{f}_{tot}.
\end{equation}
In the Coulomb gauge we have $\nabla \cdot \mathbf{A}=0$ and
furthermore let's suppose that we have a potential flow with
$\mathbf{A}=\nabla \chi$, altogether with $\chi$ a scalar potential. In hydrodynamics the analogue to $\chi$ is the potential of velocity
$\Phi$. Hence, let's suppose instead that our physical system is
submitted to an external massic force deriving from a potential $u$,
$\mathbf{f}^{ext}=-\nabla u$. Of course, this supposition is similar
to the supposition taken with Eq.~\ref{eq12a}. Note that we make use
of the Amp\`{e}re equation $\nabla \times
\mathbf{H}=\mathbf{J}=\rho_c \mathbf{v}$ (omitting the displacement
current term $\partial \epsilon\mathbf{E}/\partial t$) in order to put the term
$\mathbf{J} \times \mathbf{B}$ under the form:
\begin{equation}\label{eq12b}
[\mathbf{J} \times \mathbf{B}] = \left( \mathbf{B} \cdot
\nabla \right)\frac{\mathbf{B}}{\mu} - \frac{1}{2}\nabla \left( \frac{B^2}{\mu} \right).
\end{equation}
It can be shown using standard methods (e.g. Ref.~\cite{Landau_01}) that whenever we
multiply internally both members of Eq.~\ref{eq12a} by the
elementary displacement vector $d\mathbf{r}$ directed along the
streamline, it results into the Lagrange's integral:
\begin{equation}\label{eq12c}
\rho_c \frac{\partial \chi}{\partial t} + \frac{B^2}{2 \mu} + \rho_c (\mathbf{v}
\cdot \mathbf{A}) + \rho_c \phi + u = C(t).
\end{equation}
Here, the function $C(t)$ depends only of time. Although the first term on the right-hand-side of Eq.~\ref{eq12b} plays an important role whenever there is bending and parallel compression of the magnetic field lines we can assume here approximately straight and parallel field lines, in which case it vanishes.

Using now Eq.~\ref{eq12b} and considering that
$u=\frac{U}{\rho_m}=p+\frac{1}{2}\rho_m v^2$ with $\rho_m=mn$
denoting the mass density, we can rewrite Eq.~\ref{eq12c} under the
form:
\begin{equation}\label{eq12cc}
\rho_c \frac{\partial \chi}{\partial t} + (\mathbf{J} \cdot
\mathbf{A}) + \frac{B^2}{2 \mu} + \rho_c \phi + p +
\frac{1}{2}\rho_m v^2=C(t).
\end{equation}
This is fundamentally the law of conservation of energy, and it
states that the energy of matter plus the energy of this ``fictitious"
fluid (carrying electromagnetic fields) is constant along a
streamline. Hence, the nature of the fluid can enter through this
$u$ function, which means here the internal energy per unit mass.
The Bernoulli's integral can also be obtained in the presence of a
B-field for a particle of fluid flowing along the line of current,
since then $\partial \mathbf{A}/
\partial t=0$. Envisaging the permanent flux of a given fluid, we
aim now to integrate along a given streamline. If we do the inner
product of Eq.~\ref{eq12c} by $d\mathbf{r}$, the differential
element of a line of current it is easily obtained:
\begin{equation}\label{eq12d}
p + \rho_c \phi + \frac{B^2}{2 \mu} + \frac{1}{2}\rho_m v^2 +
(\mathbf{J} \cdot \mathbf{A})=C.
\end{equation}
$C$ (in SI units J$/$$m^3$) remains constant as long as we stand
over the line of current, changing value when we change to another
current of line. Besides the contribution of the electric potential
and the magnetic pressure terms, Eq.~\ref{eq12d} shows the
contribution of something new - a pressure term associated to the
product of the electric current density by the vector potential. This new term represents the onset of a new kind of interaction through the
agency of the vector potential, that is, by the intermediary of
fields with a bigger range of action (decaying like $1/r$). Also, it shows that the effective total pressure of the fluid has the contribution of a another kind of fluid with energy $(\mathbf{J} \cdot \mathbf{A})$ per unit volume.

Onoochin {\it et al}~\cite{Onoochin} arrived to a similar conclusion, that an
additional magnetic force to the Lorentz force should exist similar
to a magnetic pressure. It should also be related to the longitudinal forces claimed to exist~\cite{Nasilowski1,Graneau,NGraneau_01,Cavalleri_03,Graneau_01} and could introduce an electromagnetic lag in the AB effect~\cite{Boyer_73}.

The result embedded into Eq.~\ref{eq12d} can be applied to
magnetocumulative generators, devices that generate electromotive
forces through magnetic compression. It is worth to remark that
although the value of the vector potential $\mathbf{A}$ at a given
point of the physical space is not gauge invariant, the coupling
term $J_{\mu} A^{\mu}$ is gauge invariant~\cite{Casher_93}.

We may also notice that the nature of the fluid on which the
electromotive force is acting in Eq.~\ref{eq12a} does not play any
role. If the fluid is a good conductor of electricity the resulting
equations are those of the magnetohydrodynamics. Otherwise, the above results have a quite general nature.

The above integral of Eq.~\ref{eq12d} helps to mastery the principle
of producing electric fields of required configuration in the
plasma~\cite{Rickard_06}, understanding the anomalous diffusion of charged particles in magnetized plasmas~\cite{Pinheiro3}, facilitating the development of high-current accelerators, plasma-optical systems and thermonuclear devices.

\subsection{Magnus's force and Joukowski's theorem}

The Magnus force is the lift force on a cylinder moving through a
fluid when there is a net circulation of fluid around the cylinder.
Eq.~\ref{eq6} allows us to obtain in a different way the Magnus force
or the Joukowski's formula. In addition, remark that the second term of
Eq.~\ref{eq6} is similar to the Coriolis (gyroscopic) force. To calculate the
total (resultant) force acting on a given body with surface $S$ we
need to integrate all over the surface:
\begin{equation}\label{eq13}
\mathbf{F} = - \int\int\int_V \rho [\mathbf{\omega} \times
\mathbf{v}]dS dl.
\end{equation}
Here, the elementary volume is supposed to have cylindrical symmetry, such as $dV=dS.dl$. We consider the fluid
incompressible, assumptions usually used to demonstrate Joukowski's
theorem. If instead we are interested in the force per unit of
transversal length, we have to integrate to obtain
\begin{equation}\label{eq14}
\frac{|d\mathbf{F}|}{dl} = -\rho \int\int_S |[\mathbf{\omega} \times
\mathbf{v}].\mathbf{n}dS|=-\rho | [\mathbf{\omega} \times \int\int_S
\mathbf{v} \cdot \mathbf{n}dS]| = -\rho |[\mathbf{\Gamma_H}
\times \mathbf{v_0}]|=-\rho \Gamma_H v_0,
\end{equation}
where $v_0$ is the fluid velocity at infinity and where we have taken the
unit vector $\mathbf{n}$ along the outer normal to the circulation
around the contour $\gamma$, such as $\mathbf{\Gamma}=\Gamma
\mathbf{n}$.

\subsection{Force exerted by a charged particle on a body carrying current}

We intend to apply the former equations to calculate the angular velocity that an ideal solenoid carrying flux $\Phi_B$ would
communicate to an electrically charged particle of mass $m$ and charge $q$, placed at a distance $r$ from the solenoidal axis, in an azimuthal trajectory with velocity $V_{\theta}$. From Eq.~\ref{eq3}, we obtain $v_{\theta}=\Gamma/2 \pi r$, while from Eq.~\ref{eq4}, we have
$v_{\theta}=\Phi_B/2 \pi r$~\cite{Trammel}.
The canonical momentum can be written immediately, containing the particle mechanical momentum plus the ``field" (or ``fluid") momentum:
\begin{equation}\label{eq15}
\mathbf{P} = m\mathbf{V}_{\theta} + \frac{q}{c}\frac{\Phi_B}{2 \pi
r}\mathbf{u}_{\theta}=const.
\end{equation}
Therefore, the ``fluidic electrodynamics" treatment is rewarding, since considering the second term of Eq.~\ref{eq6}, we realize that the
non-null term is $\partial \mathbf{u}/\partial t$ and, accordingly, we obtain the force to which is submitted the charged particle (e.g., Ref.~\cite{Trammel,Calkin}):
\begin{equation}\label{eq16}
F_p=-q \frac{\partial \mathbf{A}}{\partial t}= - \frac{q}{2 \pi rc}
\dot{\Phi}_B.
\end{equation}

\subsection{Force between two parallel currents of ``fluid"}\label{secd}

The calculation of the force exerted between two parallel currents is a standard application of electromagnetic theory, hereby
applied with the help of our formalism. When considering our Eq.~\ref{eq6}, we realize that only the term of Coriolis has any relevance to the
problem. Hence, let us consider two filamentary vortex distant a part of $r$. The elementary force that filament (1) would exert on
filament (2) is given by:
\begin{equation}\label{eq17}
d\mathbf{F}_{2(1)}=- dm_2 [\mathbf{\omega} \times \mathbf{v}_1],
\end{equation}
where $dm_2$ denotes the element of mass of a given ``kind of flux" over which the given force is exerted. Considering the
element of mass in the elementary volume $dV=dS.dl$ around the filamentary axis (2), we can easily obtain
\begin{equation}\label{eq18}
\frac{d|\mathbf{F}_{2(1)}|}{dl_2}=- \rho \left|\int_{S_2}
[\mathbf{\omega_2} \times \mathbf{v_1}].d\mathbf{S_2} \right|,
\end{equation}
since $dm_2=\rho dS_2.dl_2$. We recall that the velocity induced
around an infinite filamentary vortex (2) $\mathbf{\omega_2}$ is
given by $v_2=-\Gamma_2/2 \pi r$ (e.g., Ref.~\cite{Prandtl}). It leads us to:
\begin{equation}\label{eq19}
\frac{d|\mathbf{F}_{2(1)}|}{dl_2} = -\rho \frac{\Gamma_1}{2\pi r}\int_{S_2}
(\mathbf{\omega_2} \cdot d\mathbf{S_2})=-\rho \frac{\Gamma_1
\Gamma_2}{2 \pi r},
\end{equation}
that we can rewrite in the form:
\begin{equation}\label{eq20}
\frac{d\mathbf{F}_{2(1)}}{dl_2}=-\rho \frac{\Gamma_1 \Gamma_2}{2 \pi
r} \mathbf{u_x}.
\end{equation}
Considering the orientation of the vorticity, we have an attractive
or repulsive force between the two filamentary vortex. Using the
analogies shown in Table~\ref{table1} it is possible to establish
the electromagnetic force between two current-carrying wires:
\begin{equation}\label{eq20a}
\frac{d\mathbf{F}_{2(1)}}{dl_2}=-\mu_0 \frac{i_1 i_2}{4 \pi
r}\mathbf{u_x}.
\end{equation}
We can use Eq.~\ref{eq12d} to obtain the pressure field around the
filaments. When considering electrical wires
acting over the ``electric fluid" in a vacuum (with no matter motion),
we obtain the scalar pressure at a given point $\gamma$ of the
streamline:
\begin{equation}\label{eq21}
p(\gamma) = p_0 - (\mathbf{J}_2 \cdot \mathbf{A}_1).
\end{equation}
Eq.~\ref{eq21} contains an interaction term of pressure, where $J_2$ is the current density in the second wire (or, in the analogue filamentary vortex) and $A_1$ is the vector potential in the first wire.
When both currents are parallel we have a pressure decrease of the
``electric fluid" in the region between the two parallel wires,
resulting thus into attraction, while when the currents are
anti-parallel repulsion will prevail. To translate to the isomorphic
relationship in the framework of hydrodynamics, we need to notice that
Eq.~\ref{eq21} involves the current {\it inside} the conductor, that
is, the vortex core, and therefore we need to use instead the expression of the speed $v=\Gamma r/(2 \pi R^2)$, where $R$ is the vortex (tube)
radius. Hence, we need to use Eq.~\ref{eq17} integrating through the
surface $S$ to obtain the pressure at the center $r=0$ of
the vortex of radius $R$ (e.g., Ref.~\cite{Prandtl}):
\begin{equation}\label{eq22}
p=p_0 -\rho\frac{\Gamma_1 \Gamma_2}{4 \pi^2 R^2}.
\end{equation}
Clearly and systematically, we see that this program of fluidic
electrodynamics leads us to a new electrodynamic force, also
discussed in Onoochin {\it et al.}~\cite{Onoochin} through a different
method.

\section{Conclusion}

The new methodological approach provided by the ``fluidic
electrodynamics" approach allows a fast transposition from electromagnetism to
fluid dynamics, and vice-versa. Under the practical viewpoint the
application of analogies allowed us to arrive, in what respect the
main objective of our research, to conclusions which are more
difficult to obtain by other methods. In addition, we remark that the proposed fluidic
electrodynamics framework described in this paper allows one to study the dynamics of any kind of ``fluid" (e.g., electro-tonic state - EM phenomena;
ordinary fluids - hydrodynamic phenomena). In such a framework, a new electrodynamic force equation was obtained with a new interaction term which represents a longitudinal force exerted among different elements of a circuit. The isomorphism between EM-fields and fluids become entirely symmetrical. The idea of longitudinal forces in electrodynamics was introduced by Amp\`{e}re and verified by himself and de la Rive with an experiment done in 1882, where a hairpin was propelled along two troughs of liquid mercury due to the longitudinal repulsion (e.g. Ref.~\cite{Johansson_96}). Other experiments seem to point toward the reality of this force, among them, the Nasilowski's wire fragmentation experiment~\cite{Nasilowski1}. As it is still going on an electrodynamic force law controversy~\cite{NGraneau_01,Cavalleri_03,Graneau_01} we believe that the theoretical frame offered in this paper reinforces the necessity to consider this kind of force as a real one.

However, this electrodynamic force equation (with a new term beyond the standard Lorentz equation) become galilean invariant. It was shown, however, that classical electrodynamics can be formulated consistently with Galilean transformations~\cite{Gomberoff_69}, in particular, through an appropriated redefinition of the fields constitutive equations~\cite{Miller_77}.

Longitudinal forces could introduce an electromagnetic lag in the AB effect, and this could possibly mean that the AB effect is local and maybe a classical effect~\cite{Boyer_73}.

The theoretical framework offered in this paper can be applied to
magnetocumulative generators and may help to mastery the principle of producing electric fields of required configuration in plasmas,
which facilitates the development of high-current accelerators, plasma-optical systems and thermonuclear
devices.


\begin{thebibliography}{99}

\bibitem{Rousseaux_02} Germain Rousseaux and \'{E}tienne Guyon, ``A propos d' une analogie entre la m\'{e}canique des fluides et l' \'{e}lectromagnetisme", Bulletin de l'Union des Physiciens {\bf 96}, 107 (2002)

\bibitem{Liu_93} Mario Liu, ``Hydrodynamic theory of electromagnetic fields in continuous media", Phys. Rev. Lett. {\bf 70} (23) 3580
(1993)

\bibitem{Sridhar_98} S. Sridhar, ``Turbulent transport of a tracer: An electromagnetic formulation", Phys. Rev. E {\bf 58} (1) 522
(1998)

\bibitem{MartPin_08} A. A. Martins and M. J. Pinheiro, ``Inertia, electromagnetism and fluid dynamics", AIP Conf. Proc. {\bf 969}, 1154 (2008)

\bibitem{Marmanis_98} H. Marmanis, ``Analogy between the Navier-Stokes equations and Maxwell's equations: Application to turbulence", Phys. Fluids {\bf 10}(6) 1428 (1998)

\bibitem{Holland_05} Peter Holland, ``Hydrodynamic construction of the electromagnetic field", Proc. Roy. Soc. A {\bf 461} 3659
(2005)

\bibitem{Volovik}Grigory E. Volovik, {\it The Universe in a Helium
Droplet}, (Clarendon Press, Oxford, 2003)

\bibitem{Mazur_05} George Chapline and Pawel O. Mazur, ``Superfluid picture for rotating space-times", accessed in
arXiv:gr-qc/0407033 5 May 2005

\bibitem{Vinem} W. F. Vinem and J. J. Niemela, ``Quantum turbulence", J. Low Temp. Phys.
{\bf 128}(5$/$6), 167 (2002)

\bibitem{Volovik_98} G. E. Volovik, ``Simulation of quantum field theory and gravity in superfluid ${}^3$He", Low Temp. Phys. {\bf 24} (2)
127 (1998)

\bibitem{Meholic} H. D. Fronning, Jr. and Gregory V. Meholic, ``Unlabored Transitions between Subluminal and Superluminal Speeds in a Higher Dimensional Tri-space", AIP Conf. Proc. {\bf 969}, 1129 (2008)

\bibitem{Kaufman1} H. R. Kaufman, ``Fluid dynamic approach to mass of classical electron", Phys. Lett. {\bf 33A} (1) 9 (1970)

\bibitem{Kaufman2} H. R. Kaufman, ``Fluid dynamic approach to the relativistic electromagnetic field", Lett. Nuovo Cimento {\bf IV} (24) 1139 (1970)

\bibitem{Pinheiro1} Mario J. Pinheiro, Phys. Essays
                    {\bf 20}(2), 1 (2007) [http://arxix.org/abs/physics/0511103]

\bibitem{AlexPin1} Alexandre A. Martins and Mario J. Pinheiro, ``The connection between inertial forces and the vector potential", AIP Conference Proceedings {\bf 880}, 1189 (2007)

\bibitem{AlexPin2} Alexandre A. Martins and Mario J. Pinheiro, ``On the electromagnetic origin of inertia and inertial mass", Int. J. Theo. Phys. {\bf 47} (10) 2706 (2008)

\bibitem{Yang_1} A. C. T. Wu and Chen Ning Yang, ``Evolution of the concept of the vector potential in the description of fundamental interactions", International Journal of Modern Physics A {\bf 21} (16) 3235 (2006)

\bibitem{Riemann} Bernhard Riemann, {\it Mathematische Werke},
2nd Edition, edited by H. Weber (Leipzig, 1897)

\bibitem{Armstrong} H. L. Armstrong, ``Comment on moving transparent media", Am. J. Phys. {\bf 37}, 336 (1969)

\bibitem{Dirac_01} P. A. M. Dirac, ``Is there an aether?", Nature {\bf 168}, 906 (1951)

\bibitem{Dirac_51} P. A. M. Dirac, ``An extensible model of the electron", Proc. Roy. Soc. London Ser. A
{\bf 268} (1332) 57 (1962)

\bibitem{Poincare_01} Henry Poincar\'{e}, ``La theorie de Lorentz et le principe de reaction", Archives n\'{e}erlandaises des sciences exactes et
naturelles {\bf 5}, 252 (1900)

\bibitem{Graham_80} G. M. Graham and D. G. Lahoz, ``Observation of static electromagnetic angular momentum in vacua
", Nature {\bf 285} 154 (1980)

\bibitem{Leonhardt} Leonhardt, U. and Piwnicki, P., ``Optics of nonuniformly moving media", Phys. Rev. A 60, 4301 (1999).

\bibitem{Corum} T. D. Keech and J. F. Corum, ``A new derivation for the field of a time-varying charge in Einstein's theory ", Inter. Journ. Theor.
Phys. {\bf 20} (1) 63 (1981)

\bibitem{Tonomura}Tonomura, A., ``Direct observation of thitherto unobservable quantum phenomena by using electrons",
Proc. Natl. Acad. Sci. {\bf 102} (42), 14952 (2005).

\bibitem{Markevitch} Douglas Clowe, Maru\v{s}a Bradac, Anthony H. Gonzalez, Maxim Markevitch, Scott W. Randall, Christine Jones, and
Dennis Zaritsky, ``A direct empirical proof of the existence of dark matter", The Astrophysical Journal {\bf 648} L109-L113 (2006)

\bibitem{Teller} Edward Teller, ``Electromagnetism and gravitation", Proc. Natl. Acad. Sci. {\bf 74} (7), 2664 (1977)

\bibitem{Bramson} Brian Bramson, ``Do electromagnetic waves harbour gravitational waves?", Proc. R. Soc. A {\bf 462}, 1987 (2006)

\bibitem{Dimtryev_01} V. P. Dmitriyev, ``Dynamics of a particle entrained in the medium flow", [http://arxiv.org/abs/physics/0612210], accessed April 20, 2007

\bibitem{Okun} J. D. Jackson and L. B. Okun, ``Historical roots of gauge invariance
", Rev. Mod. Phys. {\bf 73} 663 (2001)

\bibitem{Guyon} Etienne Guyon, Jean-Pierre Hulin, and Luc Petit,
{\it Hydrodynamique physique} (Editions CNRS, Paris, 1991)

\bibitem{Semon} Mark D. Semon, ``Note on the analogy between inertial and electromagnetic forces", Am. J. Phys. {\bf 49} (7) 869 (1981)

\bibitem{Sivardiere} J. Sivardiere, ``On the analogy between inertial and electromagnetic forces", Eur. J. Phys. {\bf 4} 162 (1983)

\bibitem{Semon_81} Mark D. Semon and Glenn M. Schmieg, ``Note on the analogy between inertial and electromagnetic forces", Am. J. Phys.
{\bf 49} (7), 689 (1981)

\bibitem{Dempsey} Daniel F. Dempsey, ``The rotational analog for Faraday's magnetic induction law: Experiments", Am. J. Phys. {\bf 59} (11),
1008 (1991)

\bibitem{Marmanis_phd} H. Marmanis, ``Analogy between the NavierStokes equations and Maxwell's equations: Application to turbulence", Phys. Fluids {\bf 10} (6), 1428 (1998)

\bibitem{Rousseaux_07} Germain Rousseuax, Shahar Seifer, Victor
Steinberg, Alexander Weibel, ``On the Lamb vector and the hydrodynamic charge", Exp. Fluids {\bf 42} (2) 291 (2007)

\bibitem{Hunt} Bruce J. Hunt, {\it The Maxwellians}, (Cornell University Press, Ithaca, 1991)

\bibitem{Schwartz} M. Schwartz, {\it Principles of Electrodynamics},
p. 141 (Dover, New York, 1972)

\bibitem{Pinheiro3} Mario J. Pinheiro, ``Anomalous Diffusion at Edge and Core of a Magnetized Cold Plasma", J. Phys.: Conf. Ser. {\bf 71}, 012002 (2007)

\bibitem{PhippsJr} Thomas Phipps, Jr., ``On Hertzs Invariant Form of Maxwells Equations", Physics Essays {\bf 6}, 249 (1993)

\bibitem{TBoyer_06} T. H. Boyer, ``Proposed Experimental Test for the Paradoxical Forces Associated with the Aharonov-Bohm Phase Shift", Found. Phys. Lett. {\bf 19} (5), 491 (2006)

\bibitem{Caprez_07} Adam Caprez, Brett Barwick and Herman Batelaan, ``Macroscopic Test of the Aharonov-Bohm Effect", Phys. Rev. Lett. {\bf 99}, 210401 (2007)

\bibitem{TBoyer071} T. H. Boyer, ``Comment on Experiments Related to the Aharonov-Bohm Phase Shift", Found. Phys. {\bf 38}(6), 498 (2008)

\bibitem{Boyer_73} Timothy H. Boyer, ``Classical Electromagnetic Deflections and Lag Effects Associated with Quantum Interference Pattern Shifts: Considerations Related to the Aharonov-Bohm Effect", Phys. Rev. D {\bf 8} (6) 1679 (1973)

\bibitem{Murad_06} P. A. Murad, ``Closed-Form Solutions to the Transient/Steady-State Navier-Stokes Fluid Dynamic Equations", AIP Conf. Proc. {\bf 813}, 1264 (2006)

\bibitem{ESJ_97} Thomas Phipps, Jr., ``Physical Significance of the Vector Potential", Elec. Spacecraft, {\bf 26}, 20 (1997)

\bibitem{Graneau_96} P. Graneau, {\it Newtonian Electrodynamics} (World scientific, Singapore, 1996)

\bibitem{Hadronic_98} R. Angulo, O. Rodrigues and G. Spavieri, ``Does the expression for the Lorentz force need to be modified?", Hadronic J. {\bf 20} (6), 621 (1998)

\bibitem{Monstein_98} Monstein, ``Electromagnetic induction without magnetic field", Electric Spacecraft Journal, {\bf 24}, 29 (1998)

\bibitem{Faddeev} C. F\'{e}diaevski, I. Vo\"{i}tkounski, and Y.
Fadd\'{e}ev, {\it M\'{e}canique des Fluides} (Mir, Moscow, 1974) (French edition)

\bibitem{Landau_01} L. Landau and Lifschitz, {\it Mecanique des
Fluides} (Mir, Moscow, 1971) (French edition)

\bibitem{Onoochin} V. Onoochin and T. E. Phipps, Jr, ``On an Additional Magnetic Force Present in a System of Coaxial Solenoids", Electromagnetic
Phenomena {\bf 3}, No. 2 (10), 256 (2003)

\bibitem{Graneau} Peter Graneau and Neal Graneau, ``Electrodynamic force law controversy", Phys. Rev. E {\bf 63} 058601 (2001)

\bibitem{Nasilowski1} Jan Nasilowksi, {\it Unduloids and striated desintegration of exploding wires}, Vol. 3, ed. W.G.Chase and H.K. Moore, (Plenum, N.Y., 1964) pp. 295-313

\bibitem{NGraneau_01} N. Graneau, T. Phipps, D. Roscoe, ``An experimental confirmation of longitudinal electrodynamic forces", Eur. Phys. J. D {\bf 15}, 87 (2001)

\bibitem{Cavalleri_03} G. Cavalleri, E. Casaroni, E. Tonni, and G. Spavieri, ``Interpretation of the longitudinal forces detected in a recent experiment of electrodynamics", Eur. Phys. J. D {\bf 26}, 221 (2003)

\bibitem{Graneau_01} Peter Graneau, ``Electromagnetic jet-propulsion in the direction of current flow", Nature {\bf 295}, 311 (1982)

\bibitem{Casher_93} A. V. Balatsky and B. L. Altshuler, ``Persistent spin and mass currents and Aharonov-Casher effect", Phys. Rev.
Lett. {\bf 70} (11) 1678 (1993)

\bibitem{Rickard_06} Matthew Rickard, Derek Dunn-Rankin, Felix
Winberg, Fred Carlton, ``Maximizing ion-driven gas flows", J. Electrostat. {\bf 64} 368 (2006)

\bibitem{Trammel} G. T. Trammel, ``Aharonov-Bohm Paradox", Phys. Rev. {\bf 134} (5B) B1183
(1964)

\bibitem{Calkin} M. G. Calkin, ``Linear momentum of quasistatic electromagnetic fields", Am. J. Phys. {\bf 34}(10) 921 (1966)

\bibitem{Prandtl} Ludwig Prandtl and O. G. Tietjens, {\it
Fundamentals of hydro- and Aeromechanics} (Dover, NY, 1934)

\bibitem{Johansson_96} Lars Johansson, Master of Science Thesis, Lund Institute of Technology, Sweden (1996)

\bibitem{Gomberoff_69} L. Gomberoff, J. Krause, and C. A. L\'{o}pez, ``Formulation of Special Relativity by Means of Galilean Transformations", Am. J. Phys. {\bf 37} (10), 1040 (1969)

\bibitem{Miller_77} M. A. Miller, Yu. M. Sorokin, and N. S. Stepanov, ``Covariance of Maxwell equations and comparison of electrodynamic systems", Sov. Phys. Usp. {\bf 20} (3), 264 (1977)

\end{thebibliography}
\end{document}